\documentclass[11pt]{article}

\usepackage{hyperref}
\usepackage{url}
\usepackage[dvipsnames,svgnames*,x11names*]{xcolor}

\usepackage[margin=1in]{geometry}

\usepackage{amssymb,amsmath}

\hypersetup{
  colorlinks=true,
  linkcolor=blue,
  filecolor=Maroon,
  citecolor=Blue,
  urlcolor=Blue}

\usepackage{graphicx}
\usepackage{makecell}
\usepackage{amsmath}
\usepackage{url}
\usepackage{booktabs}
\usepackage{authblk}
\usepackage[numbers]{natbib}
\title{Standard and reference-based conditional mean imputation}
\author[1]{Marcel Wolbers}
\author[1]{Alessandro Noci}
\author[1]{Paul Delmar}
\author[2]{Craig Gower-Page}
\author[2]{Sean Yiu}
\author[3]{ Jonathan W. Bartlett}

\affil[1]{Data \& Statistical Sciences, Pharma Development, Roche, Basel, Switzerland}
\affil[2]{Data \& Statistical Sciences, Pharma Development, Roche, Welwyn Garden City, UK}
\affil[3]{Department of Mathematical Sciences, University of Bath, UK}

\begin{document}
\maketitle
\begin{abstract}
Clinical trials with longitudinal outcomes typically include missing data due to missed assessments or structural missingness of outcomes after intercurrent events handled with a hypothetical strategy. Approaches based on Bayesian random multiple imputation and Rubin's rules for pooling results across multiple imputed data sets are increasingly used in order to align the analysis of these trials with the targeted estimand. We propose and justify deterministic conditional mean imputation combined with the jackknife for inference as an alternative approach. The method is applicable to imputations under a missing-at-random assumption as well as for reference-based imputation approaches. In an application and a simulation study, we demonstrate that it provides consistent treatment effect estimates with the Bayesian approach and reliable frequentist inference with accurate standard error estimation and type I error control. A further advantage of the method is that it does not rely on random sampling and is therefore replicable and unaffected by Monte Carlo error.
\end{abstract}

\hypertarget{introduction}{%
\section{Introduction}\label{introduction}}

The ICH E9(R1) addendum on estimands stresses the importance of a precise description of the targeted treatment
effect reflecting the clinical question posed by the trial objectives \citep{iche9r1}. One important attribute of an estimand is a list of possible intercurrent events (ICEs), i.e.~events occurring after treatment initiation that affect either the interpretation or the existence of the measurements associated
with the clinical question of interest, and the definition of appropriate strategies to deal with ICEs. The two most relevant strategies for the purpose of this article are the hypothetical strategy and the treatment policy strategy. For the hypothetical strategy, a scenario is envisaged in which the ICE would not occur. For the treatment policy strategy, the treatment effect in the presence of the ICEs is targeted and analyses under a treatment policy strategy include all observed outcomes regardless whether the subject had an ICE or not.

The ICH E9(R1) addendum distinguishes between ICEs and missing data. However, there are many connections between them. Under the hypothetical scenario, outcome values after the ICE are not directly observable and treated using models for missing data. Under a treatment policy scenario, the outcome assessments after the ICE are observable but may be incomplete because study drop out cannot be prevented. In addition, intermittently missing outcome assessments may also occur due to missed visits.

For more than a decade, a likelihood-based mixed model for repeated measures (MMRM) approach has been the de facto standard for the primary analysis of continuous longitudinal endpoints in pharmaceutical clinical trials \citep{Mallinckrodt2008}. MMRM approaches typically model the longitudinal outcomes at scheduled visits using a multivariate normal model with the visit treated as a categorical variable in each treatment group, adjustment for baseline covariates, and an unstructured covariance matrix.  This provides valid treatment effect estimates if the missing data are missing at random (MAR) after accounting for baseline covariates, the assigned treatment group, and observed outcomes and if the complete data model is correctly specified. Such a MAR assumption is often a good starting point for implementing a hypothetical strategy \citep{Mallinckrodt2020}. However, it is typically less plausible when implementing a treatment policy strategy for an ICE such as study treatment discontinuation. For this latter setting as well as for settings which include multiple ICEs handled with different strategies, methods based on imputation of missing data may be more suitable than MMRM \citep{CroEtAlTutorial2020}. For example, it may be feasible to impute missing data after study treatment discontinuation based on subjects from the same treatment group who also discontinued treatment but continued to be followed up \citep{PolverejanDragalin2020,Guizzaro2021}. This is compatible with a more complex MAR assumption which also accounts for the subject's discontinuation status.
An alternative which is especially attractive if post-discontinuation data are sparse is to use reference-based imputation methods \citep{CarpenterEtAl2013,CroEtAlTutorial2020,Mallinckrodt2020}. These methods formalize the idea to impute missing data in the intervention group based on data from a control or reference group. A typical approach to implement these imputation methods is to use multiple imputation based on Bayesian posterior draws of model parameters combined with Rubin's rules to make inferences.

In this article, conditional mean imputation for MAR and reference-based imputation of missing data is introduced, justified, and explored in a clinical trial application and a simulation study. The proposed approach differs from the conventional multiple imputation methods in several ways. First, it estimates the parameters
of the imputation model using maximum likelihood rather than via Bayesian posterior draws \citep{vanHippelBartlett2021}. Second, it uses a single conditional mean imputation rather than random imputations. 
Third, inference is based on resampling methods, i.e.~the jackknife or the bootstrap, and targets the frequentist repeated sampling variance. In contrast, the variance estimated by Rubin's rules is approximately information-anchored, i.e.~the proportion of information lost due to missing data under MAR is approximately preserved in reference-based imputation methods, but tends to over-estimate the frequentist variance for reference-based imputation methods \citep{Seaman2014,CroEtAl2019,LiuPang2016,Tang2017,Bartlett2021}.

Unlike the conventional approach, it avoids the specification of prior distributions and the complexity of Markov chain Monte Carlo (MCMC) sampling and leads to deterministic treatment effect estimates (i.e.~they are free from Monte Carlo sampling error) and, in the case of the jackknife, also to deterministic confidence intervals and $p$-values.

\hypertarget{conditional-mean-imputation-method}{%
\section{Conditional mean imputation method}\label{conditional-mean-imputation-method}}

In brief, the procedure is implemented in four steps. First, a MMRM imputation model is fitted to the observed outcome data. Second, conditional mean imputation of missing data is performed based on the parameter estimates from the MMRM imputation model and the chosen imputation method, e.g.~reference-based imputation. Third, the completed data is analysed using a simple ANCOVA model. Finally, inference is performed based on resampling techniques. All four steps are described and justified below.

\hypertarget{alignment-between-the-estimand-and-the-missing-data-strategy}{%
\subsection{Alignment between the estimand and the missing data strategy}\label{alignment-between-the-estimand-and-the-missing-data-strategy}}

Assume that the data are from a randomized two-group trial of an intervention versus control with $n$ subjects in total and that each subject $i$ ($i=1,\ldots,n$) has $J$ scheduled follow-up visits at which the outcome of interest is assessed.
Denote the observed outcome vector of length $J$ for subject $i$ by $Y_i$ (with missing assessments coded as NA (not available)) and its non-missing and missing components by $Y_{i!}$ and $Y_{i?}$, respectively.

Missing data in $Y_i$ may occur due to missed outcome assessment or may be caused by unobserved counterfactual outcomes relevant to the chosen estimand. Observed outcome data after an ICE handled using a hypothetical strategy is not compatible with this strategy.
Therefore, we assume that all post ICE data after ICEs handled using a hypothetical strategy are already set to NA in $Y_i$. Intermittent missing data and data after such ICEs will be assumed to be MAR (conditional on baseline covariates and observed outcomes). Observed outcome data after an ICE handled using a treatment policy strategy is compatible with the chosen strategy and therefore relevant for the analysis. Such post-ICE data are increasingly systematically collected in RCTs. However, depending on the study context, it may be challenging to retain subjects in the trial after ICEs such as study treatment discontinuation and missing post-ICE data may be common. Two options may be applied for imputation of missing post-ICE data in this setting: MAR imputation after inclusion of time-varying covariates and reference-based imputation methods.

Imputation under a MAR assumption after inclusion of appropriate internal time-varying covariates associated with the ICE (e.g.~time-varying indicators of treatment compliance, discontinuation or initiation of rescue treatment) have been proposed by \citet{Guizzaro2021}. In this approach, post-ICE outcomes are included in every step of the analysis, including in the fitting of the imputation model. The approach allows that ICEs may impact post-ICE outcomes but that otherwise missingness is non-informative. The approach also assumes that deviations in outcomes after the ICE are correctly modeled by these time varying covariates and that sufficient post-ICE data are available to reliably estimate the effect of ICEs on outcome data.

Depending on the disease area and the anticipated mechanism of action of the intervention, it may be plausible to assume that subjects in the intervention group behave similarly to subjects in the control group after the ICE treatment discontinuation. In such settings, reference-based imputation methods are valuable. They are implemented by combining estimated mean trajectories (and covariance matrices) from both treatment groups when determining the imputation distribution for a subject as described in Section \ref{sec:impstep}. 

\hypertarget{imputation-model}{%
\subsection{Imputation model}\label{imputation-model}}

The purpose of the imputation model is to estimate mean trajectories and covariance matrices for each treatment group in the absence of ICEs handled using reference-based imputation methods.
Conventionally, publications on reference-based imputation methods have implicitly assumed that the corresponding post-ICE data is missing for all subjects \citep{CarpenterEtAl2013}. We will also allow the situation where post-ICE data is available for some subjects but needs to be imputed using reference-based methods for others. However, any observed data in the intervention group after ICEs for which reference-based imputation methods are specified are not compatible with the imputation model described below and they will therefore be removed and considered as missing for the purpose of estimating the imputation model, and for this purpose only. Different options are possible for the handling of observed post-ICE outcomes in the control or reference group. One possibility is to also exclude them from the imputation model. This ensures that an ICE has the same impact on the data included in the imputation model regardless whether the ICE occurred in the control or the intervention group. On the other hand, if imputations in the reference group including post-ICE data can be reasonably based on a MAR assumption without accounting for the ICE, then including post-ICE data from the control group in the imputation model may be preferable. In any case, all observed post-ICE data will be included as is in the subsequent imputation and analysis steps. For subjects who experience an ICE that is to be handled by reference-based imputation of missing data, denote the visit after which the ICE occurs by $\tilde{t}_i<J$. For all other subjects, set $\tilde{t}_i=\infty$. A subject's outcome vector after setting observed outcomes after time $\tilde{t}_i$ to \texttt{NA} is denoted as $Y'_i$ and the corresponding data vector after removal of \texttt{NA} elements as $Y'_{i!}$.

The imputation model of the longitudinal outcomes $Y'_i$ assumes that the mean structure is a linear function of covariates. At a minimum, the covariates will include the treatment group, the (categorical) visit, and treatment-by-visit interactions. Typically, other covariates including the baseline outcome will also be added. Interaction terms between the treatment group and baseline covariates as well as three-way interactions between treatment group, baseline covariates, and visit can also be included. Indeed, such interactions are recommended if they are suspected to be present based on clinical considerations \citep{Sullivan2018}.
External time-varying covariates (e.g.~calendar time of the visit) as well as internal time-varying covariates (e.g.~time-varying indicators of treatment discontinuation or initiation of rescue treatment) may in principle also be included if indicated \citep{Guizzaro2021}. Missing covariate values are not allowed. This means that the values of time-varying covariates must be non-missing at every visit regardless of whether the outcome is measured or missing.

Denote the $J\times p$ design matrix for subject $i$ by $X_i$ and the same matrix after removal of rows corresponding to missing outcomes in $Y'_{i!}$ by $X'_{i!}$.
Here $p$ is the number of parameters in the mean structure of the model for the elements of $Y'_{i!}$.
The MMRM imputation model for the observed outcomes is defined as:
\[ Y'_{i!} = X'_{i!}\beta + \epsilon_{i!} \mbox{ with } \epsilon_{i!}\sim N(0,\Sigma_{i!!})\]
where $\beta$ is the vector of regression coefficients and $\Sigma_{i!!}$ is a covariance matrix which is obtained from the complete-data $J\times J$-covariance matrix $\Sigma$ by omitting rows and columns corresponding to missing outcome assessments for subject $i$. Typically, a common unstructured covariance matrix across treatment groups is assumed for $\Sigma$. However, separate covariate structures per treatment group are also possible and recommended in case heteroscedasticity between treatment groups is suspected. 

\hypertarget{sec:impstep}{%
\subsection{Conditional mean imputation step}\label{sec:impstep}}

Conditional mean imputation is based on either maximum likelihood or restricted maximum likelihood (REML) estimates of the regression coefficients and covariance matrices from the imputation model which are denoted by $\hat{\beta}$ and $\hat{\Sigma}$. As demonstrated by others, ML-based imputation is often more efficient and avoids the specification of prior distributions and the usage of Markov chain Monte Carlo (MCMC) methods which are required for imputation methods based on Bayesian posterior draws of imputation parameters \citep{Wang1998,vanHippelBartlett2021}.

In order to impute missing data for a subject, the marginal imputation distribution is first defined. For MAR imputation, the marginal imputation distribution for subject $i$ is a multivariate normal distribution with mean $\tilde{\mu}_i=X_i\hat{\beta}$ and covariance $\tilde{\Sigma}_i=\hat{\Sigma}$. For subjects randomized to the control group, the same marginal imputation distribution as for MAR imputation is also used for reference-based imputation methods. For a subject $i$ randomized to the intervention group for whom a reference-based imputation method is used after visit $\tilde{t}_i<J$, denote the predicted mean based on the assigned treatment as $\mu_i=X_i\hat{\beta}$. Denote the corresponding design matrix assuming that (counter to fact) the subject had been randomized to the control group (but using the subject's observed baseline characteristics and time-varying covariates) by $X_{i,ref}$ and denote the corresponding predicted mean from the imputation model by $\mu_{ref,i}=X_{i,ref}\hat{\beta}$. Then, the mean of the marginal imputation distribution depends on the chosen reference-based imputation method \citep[Section 4.3]{CarpenterEtAl2013}:

\begin{enumerate}
\item
  Copy reference (CR): $\tilde{\mu}_i = \mu_{ref,i}$.
\item
  Jump to reference (J2R): $\tilde{\mu}_i = (\mu_i[1], \dots, \mu_i[\tilde{t}_i], \mu_{ref,i}[\tilde{t}_i+1], \dots, \mu_{ref,i}[J] )^T$.
\item
  Copy increments in reference (CIR): $\tilde{\mu}_i = (\mu_i[1], \dots, \mu_i[\tilde{t}_i], \mu_i[\tilde{t}_i] + (\mu_{ref,i}[\tilde{t}_i+1] - \mu_{ref,i}[\tilde{t}_i)], \dots,$ $\mu_i[\tilde{t}_i]+(\mu_{ref,i}[J] - \mu_{ref,i}[\tilde{t}_i]))^T$.
\end{enumerate}

If the same covariance matrix is assumed for both treatment groups, then the covariance matrix of the marginal imputation distribution for reference-based imputation methods is also equal to $\tilde{\Sigma}_i=\hat{\Sigma}$. Otherwise, the relevant covariance matrix is derived based on the estimated covariance matrices from the intervention and control groups as described in \citet[Section 4.3]{CarpenterEtAl2013}.

Intuitively, the marginal imputation distribution under CR assumes that subjects who discontinue the intervention did not get any benefit from the intervention, J2R assumes that they did get benefit but that the benefit is lost immediately after discontinuation, whereas CIR assumes that the benefit accrued up to discontinuation is retained but that there is no additional residual benefit after the ICE. 
In practice, the most conservative imputation method is J2R while CIR is the least conservative.
J2R and CIR may be seen as relatively conservative imputation strategies for a symptomatic or a disease-modifying intervention, respectively. A more general family of reference-based imputation methods has been presented by \citet{White2020causal}.

Imputation of $Y_i$ is then according to the distribution of $Y_{i?}$ conditional on $Y_{i!}$ which is a multivariate normal distribution with mean $E(Y_{i?}\vert Y_{i!})=\tilde{\mu}_{i?} + \tilde{\Sigma}_{i?!} \tilde{\Sigma}_{i!!}^{-1} (Y_{i!} - \tilde{\mu}_{i!})$ where the subscripts ``?'' and ``!'' denote the selection of elements of $\tilde{\mu}_{i}$ and rows and columns of $\tilde{\Sigma}_{i}$ corresponding to missing and observed outcomes of $Y_i$, respectively.
Note that $Y_{i!}$ includes observed post-ICE data, hence they are being conditioned on in the construction of the imputed data. Rather than performing multiple random imputations of the missing data from the conditional distribution of $Y_{i?}\vert Y_{i!}$, we perform a single deterministic conditional mean imputation, i.e.~we impute $Y_{i?}$ by $E(Y_{i?}\vert Y_{i!})$. As demonstrated in the next section, analyzing the deterministically imputed data set using conditional mean imputation with an ANCOVA model is equivalent to performing infinitely many random imputations, analyzing each randomly imputed data set using ANCOVA, and then pooling the resulting treatment effect estimates.

\hypertarget{analysis-step}{%
\subsection{Analysis step}\label{analysis-step}}

The imputed data set using conditional mean imputation (as described above) can be analyzed using an analysis of covariance (ANCOVA) model with the outcome (or the change in the outcome from baseline) at a specific visit $j$ as the dependent variable, the assigned treatment group as the primary covariate and, typically, adjustment for the same baseline covariates as for the imputation model. The treatment effect estimator based on conditional mean imputation is then the regression coefficient corresponding to the treatment assignment and is denoted by $\hat{\theta}_{CMI}$. In many trials, the outcome at the final visit $J$ will be the primary outcome, but the method can also be applied to outcomes from another visit.

This treatment effect estimator is generated conditional on the REML estimate of the imputation model whereas reference-based imputation methods typically rely on Bayesian posterior draws from the imputation model \citep{CarpenterEtAl2013}. As demonstrated in \citet{vanHippelBartlett2021}, point estimates based on REML estimation yield slightly more efficient point estimates than methods based on Bayesian posterior draws.

\citet{vanHippelBartlett2021} further proposed to perform $M$ random imputations based on the REML estimate, analyze them, and use the average of the treatment effect estimates across random imputations as the overall treatment effect estimator. We denote the treatment effect from a single randomly imputed data set $m$ as $\hat{\theta}_{SI,m}$ and the overall average treatment effect as $\hat{\theta}_{MI,M}=\frac{1}{M}\sum_{m=1}^M \hat{\theta}_{SI,m}$. In contrast, our estimator $\hat{\theta}_{CMI}$ is based on a single imputed data set using conditional mean imputation.

However, in our particular setting of a continuous outcome analysed by ANCOVA, it is easy to show that the proposed estimator $\hat{\theta}_{CMI}$ is identical to the pooled treatment effect under multiple random imputation with an infinite number of imputations or, more precisely, that
\begin{equation} \label{eq:MiCmiConnection} 
\hat{\theta}_{MI,M} \to \hat{\theta}_{CMI} \mbox{  as  } M\to\infty \mbox{ (a.s.)}
\end{equation}
Thus, in this particular modeling setup, the estimator based on conditional mean imputation corresponds to a computationally efficient implementation of the proposal by \citet{vanHippelBartlett2021}.

To prove this, let $Z_m$ and $Z_{CMI}$ correspond to the imputed outcome vector from the visit relevant to the primary endpoint for all subjects based on the $m^{th}$ randomly imputed data set and conditional mean imputation, respectively. Then
\[\sum_{m=1}^M \frac{1}{M}Z_m\to Z_{CMI} \mbox{  as  } M\to\infty \mbox{  (a.s.)  }\]
by the strong law of large numbers (and because the randomly imputed outcomes $Z_m$ are realizations of the conditional mean imputation plus mean zero noise).
Let $X$ be the design matrix for the ANCOVA analyis model (which is identical for all multiple imputed data sets).
Denote the vector of estimated regression coefficients from the ANCOVA model by $\hat{\beta}_{SI,m}$ and $\hat{\beta}_{CMI}$ for the $m^{th}$ randomly imputed data sets and the conditional mean imputed data set, respectively.
Then

\[ \hat{\beta}_{MI,M}=\frac{1}{M}\sum_{m=1}^M \hat{\beta}_{SI,m} = \frac{1}{M} \sum_{m=1}^M (X^t X)^{-1}X^t Z_m = (X^t X)^{-1}X^t \sum_{m=1}^M \frac{1}{M}Z_m \to (X^t X)^{-1}X^t Z_{CMI}= \hat{\beta}_{CMI}\]\\
as $M\to\infty$ (a.s.).
This proves \eqref{eq:MiCmiConnection} because the treatment effect estimators $\hat{\theta}_{MI,M}$ and $\hat{\theta}_{CMI}$ correspond to the same component from the regression coefficient vectors $\hat{\beta}_{SI,M}$ and $\hat{\beta}_{CMI}$, respectively.

The proof relies on the fact that the ANCOVA estimator is a linear function of the outcome vector. For complete data, the ANCOVA estimator leads to identical parameter estimates as an MMRM model of all longitudinal outcomes with an arbitrary common covariance structure across treatment groups if treatment-by-visit interactions as well as covariate-by-visit-interactions are included in the analysis model for all covariates \citep[p.~197]{Amemiya1985}. Hence, the same proof also applies to such MMRM models. We expect that conditional mean imputation is also valid if a general MMRM model is used for the analysis but more involved argument would be required to formally justify this.

\hypertarget{resampling-based-inference-step}{%
\subsection{Resampling-based inference step}\label{resampling-based-inference-step}}

The most frequently used method for inference with multiple imputed data is Rubin's variance formula. However, Rubin's formula is not applicable to imputations methods generated conditional on the REML estimate \citep{vanHippelBartlett2021}. Rather than deriving complex analytic formulas for inference in our setting, we propose to use resampling methods such as the jackknife and the bootstrap \citep{EfronTibs1994}. Importantly, these methods require repeating all steps of the imputation procedure (i.e. imputation model fitting, conditional mean imputation, and analysis step) on each of the resampled data sets.

The jackknife requires $n$ estimates of the treatment effect based on data with one subject removed from the original data set each 
time. Denote the treatment effect estimate based on omission of subject $i$ by $\hat{\theta}_{(-i)}$ ($i=1,\ldots,n$).
Then, the jackknife standard error is defined as
\[\hat{se}_{jack}=\left[\frac{(n-1)}{n}\cdot\sum_{i=1}^{n} (\hat{\theta}_{(-i)}-\bar{\theta}_{(.)})^2\right]^{1/2}\]
where $\bar{\theta}_{(.)}$ denotes the mean of all jackknife estimates. Tests of the null hypothesis $H_0: \theta=\theta_0$ (and corresponding confidence intervals) are then based on the $Z$-score $Z=(\hat{\theta}-\theta_0)/\hat{se}_{jack}$ using a normal approximation.

Consistency of the jackknife method for variance estimation has been demonstrated for REML estimators in linear mixed effects models \citep[Chapter 1.4.4]{Jiang2007} and for M estimators \citep[Example 2.5]{Shao1995jackknife} under suitable regularity conditions.
As described, the conditional mean imputation is a simple transformation of the REML estimator of the imputation model, the observed data, and missing data indicators, and the ANCOVA analysis model is a linear transformation of the imputed outcomes.
Therefore, our procedure can also be represented as an M-estimator, i.e. as a stacked set of estimating equations for estimating the variance components and parameters indexing the mean function of the MMRM model and those for the analysis model. 

Bootstrap methods can be used as an alternative to the jackknife and have already been recommended for inference for multiple imputation \citep{vanHippelBartlett2021,BartlettHughes2020}. In most trials, the sample size $n$ should be sufficiently large to ensure that bootstrap standard errors and normal approximations are reliable. An advantage of the bootstrap over the jackknife is that if there are concerns about the adequacy of the normal approximation, then more precise bootstrap confidence intervals (e.g.~percentile or BCa intervals) and tests can be used for inference instead \citep{EfronTibs1994}. The disadvantages of the bootstrap are that the inference is no longer deterministic and that it is more computationally intensive as the required number of bootstrap samples $B$ is typically much larger than the sample size $n$. As a minimum, $B=999$ is recommended \citep[page 202]{DavisonHinkley1997} and, as described in the Appendix, substantially higher values may be required to obtain accurate $p$-values which are minimally affected by Monte Carlo error.

\hypertarget{application}{%
\section{Application}\label{application}}

We applied the proposed methods to a publicly available example data set from an antidepressant clinical trial of an active drug versus placebo which was also used in \citet{Mallinckrodt2013} and \citet{Tang2017}. The relevant endpoint is the Hamilton 17-item rating scale for depression (HAMD17) which was assessed at baseline and at weeks 1, 2, 4, and 6. Study drug discontinuation occurred in 24\% (20/84) for the active drug and 26\% (23/88) for placebo. All data after study drug discontinuation were missing and there was a single additional intermittent missing observation. The imputation model had the mean change from baseline in the HAMD17 score as the outcome, included the treatment group, the (categorical) visit, treatment-by-visit interactions, the baseline HAMD17 score, and baseline HAMD17-by-visit interactions as covariates, and assumed a common unstructured covariance matrix in both groups.
The data set included one intermittent missing value which was imputed under a MAR assumption. For imputing missing data after study drug discontinuation, we explored MAR and reference-based imputation methods. The ANCOVA analysis model also adjusted for the baseline HAMD17 value. The results for conventional Bayesian multiple imputation combined with Rubin's rules (with $M$=1,000 imputations) and conditional mean imputation using jackknife or bootstrap (with $B$=10,000 bootstrap samples) standard errors, respectively, are reported in Table 1.

\begin{table}

\caption{\label{tab:unnamed-chunk-2}Results for the data set from an antidepressant clinical trial.}
\centering
\begin{tabular}[t]{lllllll}
\toprule
\multicolumn{1}{c}{} & \multicolumn{1}{c}{} & \multicolumn{2}{c}{LS mean changes} & \multicolumn{1}{c}{} & \multicolumn{1}{c}{} & \multicolumn{1}{c}{} \\
\cmidrule(l{3pt}r{3pt}){3-4}
Imputation & Method & Drug & Placebo & Difference & Standard error & Two-sided $p$-value\\
\midrule
MAR & Bayes + Rubin & -7.639 & -4.837 & 2.803 & 1.115 & 0.013\\
 & Conditional mean & -7.636 & -4.835 & 2.802 &  & \\
 & - Jackknife &  &  &  & 1.107 & 0.011\\
 & - Bootstrap &  &  &  & 1.090 & 0.010 \vspace{.1in} \\
 
J2R & Bayes + Rubin & -6.961 & -4.839 & 2.122 & 1.122 & 0.060\\
 & Conditional mean & -6.965 & -4.839 & 2.126 &  & \\
 & - Jackknife &  &  &  & 0.858 & 0.013\\
 & - Bootstrap &  &  &  & 0.846 & 0.012 \vspace{.1in} \\

CR & Bayes + Rubin & -7.212 & -4.849 & 2.363 & 1.104 & 0.034\\
 & Conditional mean & -7.207 & -4.836 & 2.371 &  & \\
 & - Jackknife &  &  &  & 0.981 & 0.016\\
 & - Bootstrap &  &  &  & 0.968 & 0.014 \vspace{.1in} \\

CIR & Bayes + Rubin & -7.289 & -4.838 & 2.451 & 1.104 & 0.028\\
 & Conditional mean & -7.284 & -4.835 & 2.449 &  & \\
 & - Jackknife &  &  &  & 1.001 & 0.014\\
 & - Bootstrap &  &  &  & 0.986 & 0.013\\
\bottomrule
\end{tabular}
\end{table}

As shown in Table 1, conventional Bayesian multiple imputation methods and conditional mean imputation result in nearly identical estimates of the treatment effect for all imputation approaches. However, for reference-based imputation methods, the resampling-based standard errors of the treatment effect from conditional mean imputations are considerably smaller than those from Rubin's rules. The reported treatment effects estimates and standard errors are very similar to those obtained in \citet{Tang2017} using a different implementation. The treatment effect estimate and the standard error for conditional mean imputation were smallest for J2R and the bootstrap produced slightly smaller standard error estimates than the jackknife for all methods. The estimated computing time to obtain all four reported treatment effect estimates and corresponding standard errors with our R software implementation were 14.48 minutes for Bayesian multiple imputation ($M=1,000$), 5.16 minutes for conditional mean imputation plus jackknifing, and 312 minutes (5.2 hours) for conditional mean imputation plus bootstrapping ($B=10,000$, i.e. on average 1.87 seconds per bootstrap sample) on a single core of a low-power laptop with an Intel Core i5-8365U processor (1.60 GHz), 16 GB RAM, and Windows 10 Enterprise.

\hypertarget{simulation-study}{%
\section{Simulation study}\label{simulation-study}}

To further explore the statistical properties of the methods, we conducted a simulation study of a trial of an active drug versus placebo with 100 subjects per group and bi-monthly visits from baseline until 12 months. The other simulation parameters were chosen as follows:

\begin{itemize}
\item
  The mean outcome trajectory in the placebo group (and in the active group under the null hypothesis) increased linearly from 50 at baseline to 60 at 12 months.
\item
  The mean outcome trajectory in the active group under the alternative hypothesis was identical to the placebo group up to month 4. From month 4 onward, the slope decreased by 50\% to 5 points/year.
\item
  The covariance structure of the baseline and follow-up values in both groups was implied by a random intercept and slope model with a standard deviation of 5 for both the intercept and the slope, and a correlation of 0.25. In addition, an independent residual error with standard deviation 2.5 was added to each assessment. This implies marginal standard deviations ranging from 5.59 at baseline to 8.29 at the last visit in both groups.
\item
  The probability of study drug discontinuation after each visit was calculated according to a logistic model which also depended on the observed outcome at that visit. Specifically, a visit-wise discontinuation probability of 1.5\% and 2.5\% in the placebo and active group, respectively, was specified if the observed outcome was 50 or lower. For outcomes above 50, the odds of discontinuation further increase by 50\% for each 10 point increase.
\item
  Study drug discontinuation had no effect on the mean trajectory in the placebo group. In the active group, subjects who discontinued followed the slope of the mean trajectory from the placebo group from that time point onward (CIR).
\item
  Study drop-out at the study drug discontinuation visit occurred with a probability of 75\% leading to missing outcome data from that time point onward.
\end{itemize}

Under the alternative, the simulation parameters imply a true treatment effect of -2.59 (7.41 vs 10.00) under a treatment policy strategy which is the target of this simulation study. The treatment discontinuation probabilities were 34\% and 24\% in the active and placebo groups, respectively, under the null hypothesis and 31\% and 24\% under the alternative hypothesis.

Reference-based imputation methods of missing data were compared to MAR imputation. Observed post-discontinuation data was included in the analysis model but excluded from the imputation model for both reference-based and MAR-based imputation methods. 
Imputation and analysis models were as described for the antidepressant clinical trial application. Results for conventional Bayesian multiple imputation combined with Rubin's rules were based on $M=100$ random imputation and bootstrap standard errors were calculated for $B$=1,000 bootstrap samples per simulation scenario.

Results of the simulations under the null hypothesis based on 100,000 simulated data sets are presented in Table 2. 
This choice of the number of simulations provides a Monte Carlo standard error for type I error estimates of approximately 0.07\%.
As expected, the mean treatment effect estimate was essentially 0 for all imputation methods and scenarios. For Bayesian multiple imputation, the observed standard deviation of treatment effect estimates across simulation runs ($SD$ $\hat{\theta}$) was substantially smaller than the mean treatment effect standard errors from Rubin's rules (mean $\hat{se}_{\hat{\theta}}$) for reference-based imputation methods. Consequently, the simulated type I error ranged from 0.92\%-2.54\% and was substantially below the nominal type I error of 5\%. For conditional mean imputation approaches, there was close agreement between the repeated sampling standard deviation $SD$ $\hat{\theta}$ and the average standard error. The simulated type I error for the jackknife ranged from 4.84\% to 4.96\% which was slightly below but very close to the targeted type I error of 5\%. Inference based on bootstrap standard errors was associated with a small type I error inflation with simulated type I error ranging from 5.10\% to 5.26\%.

It is sometimes reported that MMRM models such as our imputation model using an unstructured covariance matrix result in convergence problems \citep{Mallinckrodt2008}. In this case, a simplified correlation structure, e.g.~a Toeplitz or autoregressive structure, may be applied. In our simulations with a relatively small sample size and a substantial proportion of subjects with treatment discontinuations and missing data, the imputation model failed to converge in only 8/100,000 (0.008\%) simulated data sets and these were excluded from our summaries in Table 1. Jackknifing did not result in any additional convergence problems and convergence problems occurred only for 1/((100,000-8)*1,000) bootstrap samples. In case convergence problems occurred for a bootstrap sample, we replaced it by a new bootstrap sample which seems unproblematic due to the minimal number of observed convergence problems. 

\begin{table}

\caption{\label{tab:unnamed-chunk-3}Results from the simulation study under the null hypothesis (for a nominal two-sided significance level of 5\%).
Simulations are based on 100,000 simulated data sets which provides a Monte Carlo standard error for type I error
estimates of approximately 0.07\%.}
\centering
\begin{tabular}[t]{llrrr}
\toprule
Imputation & Parameter & Bayes+Rubin & \makecell[r]{Conditional mean\\ (jackknife)} & \makecell[r]{Conditional mean\\ (bootstrap)}\\
\midrule
MAR & Mean $\hat{\theta}$ & 0.001 & 0.001 & 0.001\\
 & SD $\hat{\theta}$ & 0.926 & 0.926 & 0.926\\
 & Mean $\hat{se}_{\hat{\theta}}$ & 0.924 & 0.932 & 0.921\\
 & Type I error & 5.00\% & 4.94\% & 5.26\%\vspace{.1in}\\

J2R & Mean $\hat{\theta}$ & -0.005 & -0.004 & -0.004\\
 & SD $\hat{\theta}$ & 0.691 & 0.690 & 0.690\\
 & Mean $\hat{se}_{\hat{\theta}}$ & 0.928 & 0.694 & 0.687\\
 & Type I error & 0.92\% & 4.84\% & 5.10\%\vspace{.1in}\\
 
CR & Mean $\hat{\theta}$ & 0.001 & 0.001 & 0.001\\
 & SD $\hat{\theta}$ & 0.803 & 0.802 & 0.802\\
 & Mean $\hat{se}_{\hat{\theta}}$ & 0.915 & 0.807 & 0.800\\
 & Type I error & 2.54\% & 4.96\% & 5.17\%\vspace{.1in}\\

CIR & Mean $\hat{\theta}$ & 0.001 & 0.001 & 0.001\\
 & SD $\hat{\theta}$ & 0.797 & 0.796 & 0.796\\
 & Mean $\hat{se}_{\hat{\theta}}$ & 0.915 & 0.801 & 0.792\\
 & Type I error & 2.45\% & 4.94\% & 5.22\%\\
\bottomrule
\end{tabular}
\end{table}

Results of the simulations under the alternative hypothesis based on 10,000 simulated data sets (providing a Monte Carlo standard error for power estimates of \textless0.5\%) are reported in Table 3. The mean treatment effect estimate under a CIR assumption was identical to the simulation truth of -2.59 for all three methods. MAR imputation gave inflated treatment effect estimates whereas CR and J2R were conservative. Rubin's rules overestimated the frequentist standard error for all reference-based methods leading to a substantial reduction in the estimated power compared to methods based on conditional mean imputation plus re-sampling. The frequentist standard error was lowest for J2R. Surprisingly, this led to a larger simulated power for J2R than for CIR or CR. Inference based on bootstrapping led to small power gains of 0.3-0.4\% across the four scenarios compared to the jackknife.

\begin{table}

\caption{\label{tab:unnamed-chunk-4}Results from the simulation study under the alternative hypothesis (for a nominal two-sided significance level of 5\%). Simulations are based on 10,000 simulated data sets which provides a Monte Carlo standard error for power estimates of \textless0.5\%.}
\centering
\begin{tabular}[t]{llrrr}
\toprule
Method & Parameter & Bayes+Rubin & \makecell[r]{Conditional mean\\ (jackknife)} & \makecell[r]{Conditional mean\\ (bootstrap)}\\
\midrule
MAR & Mean $\hat{\theta}$ & -3.142 & -3.142 & -3.142\\
 & SD $\hat{\theta}$ & 0.925 & 0.925 & 0.925\\
 & Mean $\hat{se}_{\hat{\theta}}$ & 0.927 & 0.934 & 0.923\\
 & Power & 92.12\% & 92.11\% & 92.40\%\vspace{.1in}\\
 
J2R & Mean $\hat{\theta}$ & -2.384 & -2.383 & -2.383\\
 & SD $\hat{\theta}$ & 0.740 & 0.740 & 0.740\\
 & Mean $\hat{se}_{\hat{\theta}}$ & 0.950 & 0.747 & 0.739\\
 & Power & 74.92\% & 90.35\% & 90.74\%\vspace{.1in}\\
 
CR & Mean $\hat{\theta}$ & -2.547 & -2.547 & -2.547\\
 & SD $\hat{\theta}$ & 0.827 & 0.827 & 0.827\\
 & Mean $\hat{se}_{\hat{\theta}}$ & 0.934 & 0.835 & 0.828\\
 & Power & 80.37\% & 86.73\% & 87.12\%\vspace{.1in}\\

CIR & Mean $\hat{\theta}$ & -2.591 & -2.591 & -2.591\\
 & SD $\hat{\theta}$ & 0.822 & 0.821 & 0.821\\
 & Mean $\hat{se}_{\hat{\theta}}$ & 0.933 & 0.829 & 0.820\\
 & Power & 81.98\% & 88.30\% & 88.73\%\\
\bottomrule
\end{tabular}
\end{table}

\hypertarget{discussion}{%
\section{Discussion}\label{discussion}}

We introduced and justified conditional mean imputation based on the REML estimates of the parameters of the imputation model for MAR and reference-based imputation methods with longitudinal multivariate normal outcomes. For inference, we explored the use of the jackknife and the bootstrap.
The data example and the simulations illustrated that this leads to nearly identical treatment effect estimates compared to established methods based on multiple imputation using Bayesian posterior draws form the imputation model parameters and Rubin's rules. The resampling methods produced standard errors which were consistent with the observed repeated sampling variation. The jackknife demonstrated exact protection of the type I error in simulations with a relatively low sample size ($n=100$ per group) and a substantial amount of missing data (\textgreater25\% of subjects with treatment discontinuations) whereas the bootstrap based on a normal approximation showed a slightly increased type I error rate compared to the nominal value (up to 5.26\% at a nominal 5\% significance level). The method based on Bayesian posterior draws and Rubin's rules had inflated standard errors relative to the true repeated sampling variance, very conservative observed type I error rates, and a substantially decreased statistical power for reference-based imputation methods.

Based on our simulations, we recommend the jackknife over the bootstrap for inference because it performed better in our simulation study, is typically much faster to compute than the bootstrap, and leads to deterministic inference.
In both the data example and the simulation study, jackknife standard errors tended to be slightly larger than bootstrap standard errors. This is in line with theoretical investigations which suggest that the jackknife variance tends to be biased upwards for finite sample sizes \citep{Efron1981} whereas bootstrap variances are more susceptible to downwards bias \citep[Section 3.4.3]{Shao1995jackknife}. As a consequence, the jackknife demonstrated exact type I error protection whereas the bootstrap led to slight type I error inflation.
While we haven't formally investigated this, it could be that bootstrap methods which use more accurate methods for inference than the normal approximation have an advantage over the jackknife for very small sample sizes or other settings where the normal approximation to the test statistics is in doubt.

There are several differences between the proposed conditional mean imputation approach combined with the jackknife and conventional multiple imputation approaches based on Bayesian posterior draws combined with Rubin’s rules. First, our approach leads to deterministic treatment effect estimates, standard errors, and inference. This is particularly important in a regulatory setting where it is critical to ascertain whether a calculated p-value which is close to the critical boundary of 5\% is truly below or above that threshold rather than being uncertain about this because of Monte Carlo error. Second, our approach does not rely on the specification of prior distributions, other tuning parameters (such as the number of imputations), or MCMC sampling. Third, previous publications on reference-based imputation methods have implicitly assumed that the corresponding post-ICE data is missing for all subjects whereas our implementation also covers the setting where post-ICE data is available for some subjects but requires imputation using reference-based methods for others. Fourth, our approach provides frequentist consistent estimates of the standard error for reference-based imputation methods whereas it is well known that Rubin's rules do not provide frequentist consistent estimates of the standard error for reference-based imputation methods \citep{Seaman2014,LiuPang2016,Tang2017,CroEtAl2019,Bartlett2021}. Intuitively, this occurs because reference-based imputation methods borrow information from the reference group for imputations in the intervention group. This creates a positive correlation between outcomes in the two treatment groups and leads to a reduction in the frequentist variance of the resulting treatment effect contrast which is not captured by Rubin's variance estimator. Formally, the discrepancy is due to uncongeniality between the imputation and analysis models in reference-based imputation methods \citep{Meng1994,Bartlett2021}.
\citet{CroEtAl2019} argued that Rubin's variance (which cannot be obtained from our approach) is nevertheless valid for reference-based imputation methods because it is approximately information-anchored, i.e.~the proportion of information lost due to missing data under MAR is approximately preserved in reference-based imputation methods. In contrast, several other authors have implicitly or explicitly favored inference that is correct from a frequentist repeated-sampling perspective \citep{Seaman2014,LiuPang2016,Tang2017,Bartlett2021}.
To us, information anchoring is a sensible concept for sensitivity analyses, whereas for a primary analyses, we feel that that it is more important to adhere to the principles of frequentist inference. 

The proposed approach requires resampling for inference which is generally applicable but also computationally intensive. 
However, the time-consuming resampling procedure can easily be parallelized and computing time reduced if computations are distributed over multiple cores. A further limitation is that the validity of our approach has only been demonstrated for cases where parameter estimates from the analysis model are a linear transformation of the outcome such as general linear models including ANCOVA. This hinders the generalizability of the approach to binary or count outcomes. In addition, we have not covered the case of missing baseline covariates.

Under a standard MAR-based missing data assumption (accounting for baseline covariates, the treatment group, and observed outcomes), direct likelihood approaches such as MMRM are the most established and computationally efficient approach for treatment effect estimation. However, our approach and other approaches based on missing data imputation still play an important role for implementing more complex MAR assumptions. As an example, \citet{Noci2021} implemented a treatment policy strategy by including interaction terms between the treatment group and two time-varying covariates in the imputation model to account for the impact of treatment discontinuations and initiations of symptomatic treatment, respectively, on outcomes. Completed data sets based on this more complex imputation model were then analyzed using a standard ANCOVA analysis model. It may be possible to derive an equivalent treatment effect estimator (and a corresponding standard error) directly from an MMRM model with time-varying covariates but this would be conceptually less straightforward and require additional algebraic manipulations. Providing formal justification and reliable implementations of such alternative MMRM-based approaches is an area of ongoing research. 

We believe that methods based on missing data imputation are the most flexible and 
established approach to implement reference-based missing data assumptions but we 
acknowledge that related approaches based on MMRM model fitting have also been 
proposed by \citet{Mehrotra2017} and \citet{LiuPang2016}. 
Instead of using subject‐level imputation, \citet{Mehrotra2017} propose 
a simpler approach for the primary analysis in which the entire missing mean for 
the dropouts from the intervention group is explicitly replaced with the MAR‐based 
estimated overall control group mean. The resulting treatment effect estimator is 
the difference between the mean outcome of completers from the intervention group and 
the MAR‐based estimated mean outcome from the full control group, which is then 
multiplied by the proportion of completers from the intervention group. 
The mean outcomes in these two patient groups are obtained by separate MMRM models 
evaluated at the observed overall mean(s) of the baseline covariate(s) based on 
all randomized patients in the trial. Of note, \citet{Mehrotra2017}
compare two non-randomized patient groups (completers from the intervention group 
versus all subjects from the control group) and observed outcomes from 
subjects in the intervention group who drop out have no impact on the 
resulting treatment effect estimator. In case the probability of dropout 
depends only on baseline covariates but not on observed outcomes, this method 
leads to similar estimates as methods based on a jump-to-reference missing data 
assumption. However, in many clinical settings, it is plausible that subjects 
with worse observed outcome values are more likely to drop out. 
Consequently, completers in the intervention group would on average have better 
outcomes than the full intervention group. 
Under the null hypothesis of equal mean outcomes in the intervention and control group, 
the approach by \citet{Mehrotra2017} would therefore target a treatment effect 
different from zero which could also lead to type I error inflation. 
In contrast, none of the reference-based imputation methods included 
in our simulation study showed evidence of bias under the null hypothesis 
or type I error inflation even though the probability of study drug discontinuation depended on observed outcomes.
\citet{LiuPang2016} derived treatment effect estimators compatible with a reference-based imputation assumption based on parameter estimates from a standard MMRM fit to the data. However, it would be challenging to extend this approach to more general settings (e.g. allowing for different covariance structures in the two treatment groups or accounting for the inclusion of post-ICE data as proposed here).

Analyses of data with missing observations generally rely on unverifiable missing data assumptions and the assumptions for reference-based imputation methods are relatively strong \citep{White2020causal,Bartlett2021}. Informally, they assume that treatment discontinuations in the control or placebo group do not affect outcomes, i.e. that imputations under a standard MAR assumption are valid, and that subjects in the intervention group who discontinue treatment subsequently immediately follow the mean trajectories implied by the  chosen reference-based imputation assumption. However, these assumptions may not be plausible if subjects who discontinue randomized treatment are likely to also initiate rescue medication or if the effects of discontinuing the intervention treatment are not immediate. Therefore, these assumptions need to be clinically justified as appropriate or at least conservative for the considered disease area and the anticipated mechanism of action of the intervention treatment. Moreover, if sufficient post-discontinuation data is available, it is possible to check whether reference-based imputations of missing post-discontinuation data are consistent with the observed 
post-discontinuation data. For example, one could graphically examine whether there are any systematic deviations between observed outcomes and corresponding conditional mean imputations. Furthermore, appropriate sensitivity analyses should always be planned for missing data methods. While we have not explicitly discussed it, $\delta$-based imputation approaches which adjust imputed data sets prior to the analysis for sensitivity analysis purposes could also be incorporated in our framework in a straightforward manner \citep{CroEtAlTutorial2020}.

In summary, we believe that conditional mean imputation for MAR and reference-based imputation of missing data combined with the jackknife is a useful complementary approach to Bayesian multiple imputation for longitudinal multivariate normal outcomes. It provides replicable parameter estimates and standard errors and correct frequentist inference.

\hypertarget{software-implementation}{%
\section*{Software implementation}\label{software-implementation}}

All presented methods have been implemented in an R package `rbmi` which which is available from CRAN at \url{https://cran.r-project.org/web/packages/rbmi/}.

The data analysis of the antidepressant clinical trial is described in the `Quick start` vignette of the package.  
All simulation code has been uploaded to github and can be found at \url{https://github.com/nociale/Simulation_condmean_manuscript}.

\hypertarget{acknowledgement}{%
\section*{Acknowledgement}\label{acknowledgement}}

The authors thank Rachid Abbas for helpful remarks on a draft version of the manuscript.

\hypertarget{funding-and-disclosure-statement}{%
\section*{Funding and disclosure statement}\label{funding-and-disclosure-statement}}

Bartlett’s contribution was supported by the UK Medical Research Council (Grant MR/T023953/1).

Bartlett’s institution has received consultancy fees for the author's advice on statistical methodology from AstraZeneca, Bayer, Novartis, and Roche. Bartlett has received consultancy and course fees from Bayer and Roche.

\hypertarget{a-appendix-additional-details-on-bootstrap-methods}{%
\section*{A Appendix: Additional details on bootstrap methods}\label{a-appendix-additional-details-on-bootstrap-methods}}
\addcontentsline{toc}{section}{A Appendix: Additional details on bootstrap methods}

\hypertarget{a.1-bootstrap-confidence-intervals-ci-and-tests-for-the-treatment-effect}{%
\subsection*{A.1 Bootstrap confidence intervals (CI) and tests for the treatment effect}\label{a.1-bootstrap-confidence-intervals-ci-and-tests-for-the-treatment-effect}}
\addcontentsline{toc}{subsection}{A.1 Bootstrap confidence intervals (CI) and tests for the treatment effect}

A large number of bootstrap methods for CI estimation and testing have been proposed in the literature. We first discuss methods based on the bootstrap standard error and the normal approximation and percentile bootstrap methods \citep{EfronTibs1994,DavisonHinkley1997}.

Denote the treatment effect estimates from $B$ bootstrap samples by $\hat{\theta}^*_b$ ($b=1,\ldots,B$). The \emph{bootstrap standard error} $\hat{se}_{boot}$ is defined as the empirical standard deviation of the bootstrapped treatment effect estimates. The corresponding two-sided normal approximation ($1-\alpha$) CI is defined as $\hat{\theta}\pm z^{1-\alpha/2}\cdot \hat{se}_{boot}$ where $\hat{\theta}$ is the treatment effect estimate in the original data set, i.e.~$\hat{\theta}=\hat{\theta}_{CMI}$. Tests of the null hypothesis $H_0: \theta=\theta_0$ are then based on the $Z$-score $Z=(\hat{\theta}-\theta_0)/\hat{se}_{boot}$ using a standard normal approximation.

Alternatively, \emph{percentile bootstrap} methods can be used for inference: For this method, a two-sided $(1-\alpha)\%$ CI is defined as the interval from the $((B+1)\alpha/2)^{th}$ to the $((B+1)(1-\alpha/2))^{th}$ ordered bootstrap treatment effect estimate \citep[page 202f]{DavisonHinkley1997}.
Typically, $B$ is chosen such that $(B+1)\alpha/2$ is an integer value. If this is not the case, the corresponding empirical bootstrap quantiles can be calculated via interpolation. The advantage of the percentile method over the standard normal approximation is that it is also valid if the treatment effect estimator itself is not normally distributed around the population treatment effect but if a monotone transformation exists which normalizes the estimator \citep[page 173f]{EfronTibs1994}. For example, this is the case for $t$-distributed test statistics.

A \emph{one-sided} bootstrap $p$-value for a test of the null hypothesis $H_0: \theta=\theta_0$ versus the alternative $H_1:\theta > \theta_0$ can be defined via inversion of the percentile CI.
If exactly one of the bootstrap samples results in an estimate of $\theta_0$, then the $p$-value is defined as 
$\hat{p}= (\#\{\hat{\theta}^*_b < \theta_0\}+1)/(B+1)$ \citetext{\citealp[page 214-215]{EfronTibs1994}; \citealp[adding +1 to comply with the percentile bootstrap definition of][]{DavisonHinkley1997}}. Otherwise, one needs to numerically determine to which quantile of the bootstrap distribution $\theta_0$ corresponds.
The \emph{one-sided} test in the other direction can be defined in the same way, and the corresponding two-sided $p$-value is defined as 2 times the lower of the two one-sided $p$-values.

More complex bootstrap confidence intervals also exist and the BCa (bias-corrected and accelerated) CI is often recommended \citep{Carpenter2000}. This would be also possible but not trivial in our setting: Typically, $a$ is estimated based on the skewness of the score function \citep[page 205]{DavisonHinkley1997} (which is not available in closed form in our setting) or on a jackknife approximation to it \citep{Carpenter2000} (which is computationally intensive in our setting).

Permutation tests and associated bootstrap tests which draw bootstrap samples under the null hypothesis of equality between the two randomized treatment groups \citetext{\citealp[chapters 15 and 16]{EfronTibs1994}; \citealp[chapter 4]{DavisonHinkley1997}} could also be applied but have not been explored in our setting.

\hypertarget{a.2-how-many-bootstrap-samples}{%
\subsection*{A.2 How many bootstrap samples?}\label{a.2-how-many-bootstrap-samples}}
\addcontentsline{toc}{subsection}{A.2 How many bootstrap samples?}

As a general rule of thumb, \citet{DavisonHinkley1997} (pages 156 and 202) advise to use $B=999$ or more for CI estimation and testing. However, as illustrated below, substantially larger values of $B$ may be required for accurate estimation of $p$-values. Let $\hat{p}_{\infty}$ denote the one-sided $p$-value that would be obtained under the theoretical bootstrap distribution (including all possible bootstrap samples) and $\hat{p}_{B}$ the corresponding $p$-value based on $B$ bootstrap samples.
We use the following approximate distributions for $\hat{p}_{B}$:

\begin{itemize}
\item
  $\hat{p}_{B}\sim \Phi(\Phi^{-1}(\hat{p}_{\infty})/\sqrt{Z/(B-1)})$ with $Z\sim\chi^2(df=B-1)$ for the normal approximation
\item
  $\hat{p}_{B}\sim (Z+1)/(B+1)$ with $Z\sim Bin(B,\hat{p}_{\infty})$ for the inverted bootstrap percentile method
\end{itemize}

Based on these approximations, the probabilities below were evaluated:

\begin{tabular}{l|l|l|l}
\hline
\bf{Method} & \bf{B=999} & \bf{B=9,999} & \bf{B=99,999}\\
\hline
\bf{Normal approximation} &  &  & \\
\hline
95\% range for $\hat{p}_B$ if $\hat{p}_{\infty}=2.5\%$ & 2.02\% to 3.02\% & 2.34\% to 2.66\% & 2.45\% to 2.55\%\\
\hline
Probability of $\hat{p}_B\leq 2.5\%$ if $\hat{p}_{\infty}=2\%$ & 98.37\% & 100.00\% & 100.00\%\\
\hline
Probability of $\hat{p}_B > 2.5\%$ if $\hat{p}_{\infty}=3\%$ & 96.42\% & 100.00\% & 100.00\%\\
\hline
\bf{Percentiles method} &  &  & \\
\hline
95\% range for $\hat{p}_B$ if $\hat{p}_{\infty}=2.5\%$ & 1.70\% to 3.60\% &  2.21\% to 2.82\% & 2.40\% to 2.60\%\\
\hline
Probability of $\hat{p}_B\leq 2.5\%$ if $\hat{p}_{\infty}=2\%$ & 84.66\% & 99.97\% & 100.00\%\\
\hline
Probability of $\hat{p}_B > 2.5\%$ if $\hat{p}_{\infty}=3\%$ & 84.54\% & 99.88\% & 100.00\%\\
\hline
\end{tabular}

These tables demonstrate that the approximation of the bootstrap distribution with a finite number of bootstrap samples $B$ requires a larger $B$ for high accuracy of the percentile method (which is valid under weaker assumptions) than for the method based on the bootstrap standard error (which relies on a normal approximation).

\bibliography{rbmi.bib}

\end{document}